**Near-ideal selection for the Standard Genetic Code**

Michael Yarus

Department of Molecular, Cellular and Developmental Biology
University of Colorado  Boulder
Boulder, Colorado 80309-0347
Phone: +1 303 817-6018
Email: yarus@colorado.edu

**Abstract**

Evolutionary theorizing resembles building an aircraft while also piloting it; new results change the scaffold for older ideas, requiring revised strategy to remain airborne. A calculated kinetic pathway exists **(1)** that, under explicit quantitative assumptions, delivers the SGC (Standard Genetic Code). The pathway and evidence for it is summarized below, striving for a clearer, more complete account than was possible during its construction. Beginning with experimental amino acid-RNA interactions, code assignments are fused, codes divide and an early coding crescendo is tested for homogeneous assignments, which are then selected for independent survival in a near-empty biotic world. During escape from the site of origin and diaspora, a near-complete SGC becomes dominant because it supports proficient division. Crescendo, escape and diaspora together comprise a near-ideal least selection for a Standard Genetic Code that specifically served LUCA, likely a free-living anaerobic thermophilic microbe. Diaspora's selection for flexibility conceivably made the SGC's persistence across gigayears feasible.

**Discussion**

**The RNA world (2).** The most ancient **(3, 4)** coding (Fig. 1) must be irreducibly simple. Coding by reaction between bound amino acids on an RNA template can proceed using only two molecules **(5)**, an RNA template with amino acid binding sites **(6)**, and carboxyl-activated amino acids. In order to evolve, primitive encoding must also be expandable - able to produce functional peptides of varied lengths **(7)**. This enables active peptides to evolve continuously and simply, by one-amino-acid extensions. Primitive coding must join seamlessly to more modern translation, perhaps using similar molecules **(8)**. These requirements are met by direct RNA templating of the earliest encoded peptides **(9)**.

**RNA world: Cognate triplets in RNA-amino acid binding sites.** Recurrence of cognate coding triplets within amino acid binding sites on selected riboligonucleotides cannot be explained by chance. The data have been reviewed in detail **(10)** - but for example, if coding triplet sequences are reversed, 3' to 5', to yield non-cognate triplets of the same base composition, their association with binding sites disappears **(11)**. Thus, association of arbitrary triplets (even having SGC base compositions) with RNA binding sites is not observed. Alternatively, one can vary the code: when $5 \times 10^6$ codes randomized in varied ways were asked if their triplets recur in experimental amino acid sites, 99.2 to 99.5% show less association than the biological SGC **(11)**. Minor observed associations are attributable to re-creation of SGC-like structures in randomly-constructed coding tables. Thus varied tests, largely independent of assumed statistical distributions, confirm that experimental SGC coding triplets and amino acid affinities are unexpectedly associated, supporting a functional relation.

Using the experimentally-determined amino acid binding capacities of selected RNAs **(10)**, it is estimated that amino acid-RNA affinities could encode about 15 of the standard 20 amino acids **(12, 13)**. Therefore the RNA world can support an earliest era of genetic encoding, during which up to 15 amino acids could be associated with RNA triplets **(1, 13)**.

This is vital, because modern protein enzymes have been reduced to 9-15 different amino acids, while retaining both the ability to fold and to observably catalyze their reactions **(18–22)**.

Moreover, even short peptides can be catalysts. The most investigated example is Ser-His, which catalyzes the polymerization of amino acid esters into peptides **(14),** nucleotide imidazolides into dinucleotides **(15)**, and can speed fatty acid vesicle growth, via its synthetic hydrophobic peptides **(16)**. Peptides 7, 8, 9, 10, and 11 amino acids long, composed of 5 or 6 different side chains, are catalytic hydrolases **(17)**.

What simplified peptides and RNA could do together remains an urgent question. But existing observations suggest that coding employing RNA specificities alone **(6)** can encode peptides sufficiently complex to contribute to ribozyme catalysis **(23)**. Consequently, peptides and RNAs can collaborate to expand encoding during addition of the remaining 20 standard amino acids (Fig. 1, Methods). Using these estimates, some properties of the RNA world and the subsequent RiboNucleoPeptideTransition (RNPT) can be computed **(1)**.

**Expression before the DNA genome.** Ordering, controlling, and preserving expression of genetic information was presumably the advantage conferred by selection of a DNA genome. Thus, it is likely that effective expression, encoding chemical actions in nucleic acid sequences, existed before evolution of the DNA genome. Expression thus existed before the onset of Darwinian evolution (Fig. 1).

Early encoded peptide expression must be of the simplest kind. In fact, RNA-bound peptide His-Phe requires a structure with only ≈ 30% more ribonucleotides than for either His and Phe alone, so multiple proximal RNA-bound amino acids efficiently use binding ribonucleotides **(6)**. These peptide-linked amino acids are bound in sites with new structures, containing unselected cognate SGC triplets **(10)** for L-His and L-Phe. These data suggest coding by binding activated amino acids directly to an RNA template. The evolutionarily-favored peptidyl transferase itself is a simple entropic catalyst **(24, 25)**, a mechanism well-suited to ancient RNA-bound amino acids, side-by side.

**Agents of early expression.** Simple expression can also have utilized nucleotide cofactors **(26, 27)**. Cofactor homologues can be templated on ribohomopolymers **(28)**, or synthesized by ribozymes **(29)**, as well as incorporated at the 5-prime terminus of transcribed RNAs **(30, 31)**. Such cofactor-initiated RNAs still exist **(32, 33)**. In addition, cofactors are readily bound by RNAs, as in riboswitches **(34)**. Some cofactors need no host molecule, being chemically active as free molecules **(35)**.

Complex amino acid biosynthesis utilizes cofactors. So cofactors expand an RNPT amino acid repertoire. Notably, both nicotinamide (NAD, **(36)**) and pyridoxal phosphate (PLP, **(37)**) have simple, credibly prebiotic, syntheses. Given NAD and PLP catalysis, 15 of the proteinogenic 20 amino acids can be available to early microbes **(38)**. Only Val, Leu, Ile, Gly and Cys remain, all available from credible geochemistry **(39–41)**. In addition to ribonucleopeptides, it therefore seems likely that cofactors and cofactor-RNAs **(42, 43)** supplied biosynthetic amino acids to the RNPT (Fig. 1), as first envisioned by White **(26)**.

**Biology as Anthology.** Evolution can quickly unite multiple capabilities in a single organism. Combining phenotypic advances is so frequent during evolution that it deserves a name: anthology. Anthology **(44)** suggests that biology's essential achievement is merging evolutionary advances made for different reasons, in different places, or at different times.

Accordingly, mechanisms of evolutionary joining of are of great importance. Varied events in this text have this attribute. Code fusion is an example, allowing both fast access to the SGC, concurrently constructing an accurate **(45)** code. Thus the SGC is a successful anthology **(46)**.

But more or different amino acids might be encoded **(47, 48)**; and given that anthology joins several elements, there will also be varied ways to assign its codons. So, anthology also implies an intricate code distribution to be resolved.

**Evolutionary anthology: upper tails**. Complex evolutionary advances combine improvements. Perhaps many or all combinations occurred under primordial conditions. Thus, evolutionary progress is necessarily made by, in one-or-another sense, favoring the upper tail of a distribution. This complements the idea that fastest progress consists of truncating such a distribution, taking only upper tail members who exceed some threshold for a favored quality (heat tolerance, fecundity, …**(49)**. So, fitness is accurately a 'distribution fitness' **(12)**, characteristic of a distribution's upper tail.

**Anthology: least selection.** Distribution is intrinsic to a principle of least selection **(49)**. This asserts that evolution's path will be firstly, be the quickest practical route to a goal. Least selection also secondly, reaches near as possible to the goal, requiring the smallest further change. Evolution transits the distribution's tail nearest a favorable phenotype.

**Anthology: time of selection.** Upper tail selection has particular characteristics, which are therefore characteristics of the evolution of complex abilities. For an SGC example, the abundance of immediate SGC precursors is a function of time **(50)** – so selection probably occurs near a precursor peak. Given a simple unimodal kinetic evolution **(1)**, probable selection times are evident (Methods).

**Anthology: rarity.** The more complex the machinery selected, the less frequent it is likely to be. A particular favored combination can be assembled from functional parts that tend to be improbable and rare. But even if favored characters are not individually rare, small upper tail probabilities result when multiple probabilities are multiplied. Thus principal evolutionary advances are intrinsically difficult to select and explain.

**Anthology: amplification.** Because of their initial rarity, significant evolutionary advances are necessarily coupled to a mechanism for amplifying success. The genetic code is an example; SGC-like codes combining broad coding capability and accuracy are rare, appearing amidst a jumble of less competent nascent codes **(45)**.

**Anthology: no time for errors.** The well-functioning code minority in an upper tail have often gotten there quickly **(12)**. A random, quick subset has inadvertently spent no time in detours; for example, in recovery from errors. Earliest codes will be selected before systems with any other history. This is likely the reason that the fastest evolving are usually also the most accurately formed codes **(1, 51)**. Thus, correlation between evolutionary speed and accuracy arises because they are the same phenomena, viewed two different ways. SGC-like codes appear as a persistent kinetic peak of complete **(45)** codes called the crescendo (Fig. 1).

**The RiboNucleoPeptide Transition (RNPT).** In the RNPT, an initial RNA code evolves to use peptide chemistry to enhance coding by adding biosynthetic amino acids **(13)**. This era incorporates later amino acid biosynthesis, as envisioned by Wong **(52)**. The RNPT is also the principal era when code fusion effectively promotes code evolution **(1)**, enhancing code homogeneity **(45)**. In contrast, though it seems plausible that faster code divisions by more complete codes **(53)** should aid the RNPT, this does not prove to be so **(1, 51)** for codes initiated by single assignments. Instead, more efficient encoding of cell/code division becomes crucial for simply-initiated nascent codes only later (Fig. 1), during escape and diaspora **(1)**.

**RNPT: completion complications.** Special kinetic obstacles arise when a code evolves into the SGC's immediate vicinity. Here unassigned triplets (precursors for assignment) are minimal, and assigned triplets (precursors for decay) are maximized, so net triplet assignment slows. Completion complications are potentially large: the last two triplet assignments can require more than 100-fold more time than the first 20 **(12)**. This is a reason that initiation and termination codon mechanisms were fixed late, using disparate mechanisms in separated domains of life **(54)**.

**RNPT: late wobble.** Interference between code regions being assigned is a general evolutionary constraint. For example, codes are more inaccurate, the greater the number of triplets simultaneously assigned **(46)**. A specific example is wobble coding. Wobble is emulated in present calculations in a simplified form that supposes only unmodified nucleotides: G pairs with either U or C **(12)**. This assigns 2 triplets, and also deceases evolutionary accuracy **(44)**. Superwobble **(55)**, which assigns 4 triplets at a time, is yet more obstructive **(56)**.

Yet wobble coding is a near-ubiquitous feature of the SGC: 18 of 20 encoded amino acids employ it. But firstly: accurate wobble pairing requires a very particular tRNA structure **(57, 58)**. Secondly, accurate wobble employs a large allosteric ribosome structure to verify base pairing **(59)**. Thirdly, early wobble pairing hinders evolution as does any other assignment mechanism that alters multiple triplets **(46)**. Wobble obstructs the completion of the code by specifically reducing appearance of immediate precursors of the SGC **(44)** because larger assignments are more likely to conflict during code fusions. In fact, wobble becomes an increasingly serious obstacle as evolution approaches the SGC **(44)**. Thus, continuous wobble obstructs complete encoding **(12)**. These four observations suggest that wobble will appear late in SGC evolution, after most assignments are made. However, wobble's introduction can be rationalized: this work detects a time of minimal disruption due to wobble evolution, late in the RNPT **(51)**. This provides a fifth support for later wobble evolution, but simultaneously suggests how it can arise (Fig. 1, Methods).

**RNPT: terminus for freezing.** The RNPT marks the upper limit for the time Crick envisaged **(60)**, when the nascent amino acid code would "freeze" in order to preserve the functions of earlier active peptides **(61)**. Assignments presumably freeze shortly after they appear, when they have been significantly used. Thus even in the RNA world, with ≈ 15 amino acid assignments **(1)**, and certainly during the RNPT, with its multiple enhanced anabolic pathways **(62, 63)** and 20 assignments (Fig. 1), the nascent SGC is resisting fundamental alteration.

**RNPT: crescendo winnows the code, resists alternatives.** Simplest code beginnings are most probable. Thus conclusions (Fig. 1) focus on codes adding single assignments to one primordial encoded function **(13)**. Evolution of codes initiated by one assignment accelerates by fusing codes, specifically during the RNPT **(1)**.

Code fusion also has a kinetic rationale **(45)**; fusion unites separately evolved codes, providing more rapid process toward a complete SGC by joining coding progress made independently. More complete codes, closer to the SGC, are eventually selected **(1)**; this may be as simple as their more effective support for cell/code division during escape and diaspora (Fig. 1).

Code fusion can also homogenize evolving codes. Because fusions that assign different meanings to the same triplet will be less fit, ambiguous fusions can go extinct. This implies that fusion survivors display more homogeneous codes than their pre-fusion precursors (45). Descendants therefore converge toward a consensus of coding ancestors. More strikingly, and only for fusing codes, a long era with increasingly competent coding (the crescendo) exists. Code fusions present better codes sooner, are more abundant, occur in smaller populations, and persist for a longer time than excellent unfused codes. They therefore are the likely progenitors of the SGC (45).

Via the same mechanisms, coexisting independent codes, even though they may employ the same biochemistry, are discriminated against. A coexisting code which shares no assignments at all contributes only 0 to 2 assignments to the final SGC-like code **(13)** even if many random assignments potentially include it.

Code division and code fusion together effectively refine codes (51), speeding evolution and making it more accurate. During the most rapid evolution, fusing codes are almost equivalently effective (in fact, indistinguishable) when they originate from internal divisions of a single ancestor, or by independently originating-and-evolving codes (51). One can imagine a code origin suited to a poor archaic environment, where only one initial code can form (Fig. 1, Methods). If a unique mother code divides, descendants gather new evolutionary assignments, and, fusing back to the mother code, efficiently create the SGC (51). Code divisions independent of code completeness (Methods) are always the most effective, consistent with other conclusions about division, where code-completeness-dependent division is important only during diaspora **(1)**.

**SGC era: excellence in randomness.** Random sequences of chemicals undergoing random reactions host near-ideal reactions **(64)**. This may seem a surprising emphasis. But random reactions necessarily execute all options, including whatever minority are near-ideal for any particular chemistry. Notably, if evolution can capture near-ideal events, it would provide a potent route for evolutionary improvement, applicable to any pathway. Such effective capture of random near-ideal events is not only possible, but frequently expected **(65)**.

An illustration: there is a simplified form of templated RNA synthesis **(64)**, in which random nucleotide reactants first synthesize an RNA template, and then use it to instruct templated synthesis of a chemically reactive RNA product. This kind of genetic expression model imposes complex requirements, because a specific first reaction (to make template), then a second one (to make product) must occur in sequence. When replication representing this entirely random system is examined, it comes mostly from a minority of reactions in which reactants arrive in near-ideal amounts and at near-ideal times to construct a RNA replication sequence **(64)**.

**SGC era: both near-ideal and least selection.** The question has been asked: how does evolution function before the genome? The original answer was 'chance utility': that is, chance fluctuation yields an opportune moment for selection **(65)**. Here I generalize: chance utility is capture of near-ideal random events. The SGC emerges via a least selection that attains minimized time and maximized accuracy using one unified, near-ideal, random opportunity.

**SGC era: least selection.** Crescendo, escape and diaspora comprise a least selection **(49)**. Least selection implies that evolution can be facilitated either by increased speed or more accurate focus, or both. Selection of the first code fit for the world while rotating through many SGC-like codes precisely meets both speed and accuracy requirements for least selection.

**SGC era: near-ideal selection.** The emergence of a free-living code organism from a varied population is near-ideal. Code fusions rapidly construct SGC-like codes **(46, 51)**. But the same processes that construct the crescendo rapidly vary its codes **(45)**, while favoring SGC-like coding **(1)**, and disfavoring code assignments using incompatible principles **(13)**. This mixture of SGC-like codes is continuously tested by accidental escape from its origin, and possible diaspora into the surrounding world **(1)**. Escape and diaspora continue until an organism and code capable of colonizing the surroundings, perhaps progressively colonizing successive pools, arises (Fig. 1). That code is LUCA's, the escapee most competently dividing in novel environment(s) **(53)**; it later descends to all life (Fig. 1) on Earth **(1)**. The SGC-like defector and winner **(66)** can use all forms of catalysis to encode proficient division (Fig. 1), leaving behind codes limited to slower RNA-catalyzed means **(67)**. As important as this transition is, note that we are early in code history; the now-DNA-borne code still has many dramatic twists ahead **(68)**.

**SGC era: differential growth.** Escape and diaspora's selection for growth rate make even a rare code preeminent. This provides the essential amplification for a rare nascent SGC. Escape and diaspora were probably particularly effective in the relatively empty Earth of the time. Escaped microbes, with little competition, multiplied during diaspora to ultimately be dominant **(1)**. Later in this era, differing complete codes first exist and thus can compete, possibly benefiting the most communally useful of them **(69)**.

**SGC era: a stepped diaspora.** Diaspora likely required superior colonization of a series of environmental pools increasingly distant from the code's origin. A chain-of-pools diaspora is advantageous, reducing the evolution of the fully independent organism/code to smaller, more probable steps **(1)**. Fig. 1 incorporates the superiority of continuous escape, followed by diaspora in steps.

If the environment for diaspora was that for LUCA, then the code was likely formed for a thermophilic anaerobe **(63, 70)**, living on a planet we would barely recognize. Accordingly, the emerging code served a very different metabolic and ecological adaptation than today's. It is a remarkable, fundamental finding that a once-primitive SGC has survived gigayears, serving primitive unicellular microbes, complex multicellular creatures and all their intermediates.

**SGC era: distribution fitness.** Selection of individuals fit for life in a novel environment is distribution fitness **(12)**, calling for a complex, rare upper-tail behavior not previously required of microbes and codes.

**SGC era: a starting bloc.** Escape and diaspora are a starting bloc selection **(71)**, in which the first to exhibit an ability, to live freely, are particularly easily selected (Fig. 1).

**Code evolution: extraordinary choice.** Allowing partial coding, there are $20^{64}$ = 1.8 x $10^{83}$ possible genetic codes for 20 amino acids. But all life on Earth shares a unique, similar SGC **(72)**, emergent from its unthinkable abyss of choices. Choice of one option among ~ $10^{83}$ requires explanation.

Code fusion, via its ability to make codes converge on consensus assignments **(45)**, is part of the trend to code unity. Moreover, code fusion provides an arena for rejection of alternative codings, both during the most probable assignment-by assignment code evolution as well as during possible encounters with differently-evolved, near-complete codes **(13)**. In addition, the starting pool of codes itself is radically constrained: a limited number of specific affinities between primordial amino acids and primordial RNAs limits initial coding choices to a tiny fraction of those calculated from combinatorics **(10)**. Finally, escape and diaspora enforce a final profound narrowing **(69)** of coding possibilities **(1)**, taking a starting bloc's step toward the SGC among its astronomically-numbered $1.8 \times 10^{83}$ code possibilities.

**Extreme SGC versatility.** Escape and diaspora yield apparent breaks from previous genetic averages. Distributions are conditional, divergence can be rapid **(49)**. More recent Earth history, more visible to us than prebiotic times, exhibits many such events. The Great Oxygenation Event (2.4 gigayears ago) enforced swift adaptation to oxygen, poisonous to organisms thriving on a prior anaerobic Earth **(73)**. The Cambrian evolutionary 'explosion' (520 million years ago) displaced existing fauna, rapidly elaborating modern multicellular body plans **(74)**. The KT impact extinction (65.5 million years ago, **(75)**) annihilated non-avian dinosaurians and opened the Earth for new species, including mammals. And this is a small sample of relatively quick multiple changes **(76)**. The SGC accommodated them all. Diaspora plausibly enabled the extraordinary scope of Earth biota by selecting a code sufficiently versatile to succeed in varied settings.

Accordingly, Earth's evolution could have chosen differently. JBS Haldane mused: the universe is not only stranger than we imagine, but stranger than we <u>can</u> imagine. Aliens using their own conserved code will differ in unthinkable ways. If humans survive to view their reflections in an intelligent alien's visual organ, the reflector is unlikely to be an alternative ape.

**Methods**

**Evolution can be studied rigorously.** Evolution is sometimes said to be challenging because 'you can really never know how it happened' because 'there is no direct evidence'. But this prejudice is not just incorrect, it is obviously incorrect. Uncountable books are required to hold what we reliably know about things that cannot be approached directly: climatology, cosmology, ecology, geophysics, high energy physics, human history, and particularly, evolution in any but the simplest creatures. What is the resolution of this apparent incongruity?

Direct access to a subject is inessential because of an implication of Bayes' theorem **(77)**, known since the 18<sup>th</sup> century. Leaving details to the reference, Bayes says that you can plausibly and rigorously evaluate a current hypothesis in light of new data. Just multiply its present probability by the ratio: (probability of the data if the hypothesis is true)/(probability of the data in light of all possible explanations). Consequently, Bayes tells you how to strengthen or weaken any idea that makes predictions, even if you must span vast spaces and times. If earlier intermediates are also likely (say, from chemical or physical data), then Bayes directs that we explain both, and propose a pathway that joins such intermediates and the SGC. By explaining present observations better you refine your hypothesis – very much as you would using specific laboratory data. Seeking coding's origins, we rely on Bayes' assurance that we can hold course toward the SGC.

**Comparing evolutionary results.** Above, evolutionary hypotheses are compared by computing their outcomes, and comparing these to the broadly-conserved biological SGC. Outcomes are endpoints of normal kinetics, where coding events (e.g., the assignment of an amino acid to a triplet) have a set probability during a time called a passage. Passages are short, so only one evolutionary code event will occur therein **(12)**. Under these conditions, time as passages define a kinetics (called Monte Carlo Kinetics, **(51)**) in the same way as assignment of rate constants for coding events **(12)**. To accurately define mean results, many repetitions are performed: for example, in **(13)**, 1500 evolving locales (termed environments) were examined in which a total of 32,000 to 109,000 coding tables evolve to express $\geq$ 20 encoded functions. Because individual Monte Carlo Kinetics for coding tables are followed, this procedure yields distributions of evolutionary variables, as well as means. This technique was used to show that a subset of pathways to the SGC has uniquely favorable characteristics **(51)**. Programmed Monte Carlo kinetics have been described in more detail in **(1, 12, 51)**, current parameters are listed **(13)**, a numerical sample of results can be viewed **(51)**, code intermediates can be seen in standard form **(12, 13)**, and source code capable of all published calculations (Ctable25d.pas, including commented variations) is online at CERN Zenodo **(78)**.

**Distance to the SGC; least selection.** To compare evolutions, one must measure distance between varied results and the SGC. This was initially done **(12)** by defining three distances based on matching the SGC's close spacing for identical assignments, minimizing mutational distance from the SGC, and reproducing obvious SGC chemical order. However, this was later replaced with a simpler, more specific distance based on least selection **(49)**. Least selection seeks, simultaneously, fastest evolution and smallest difference from end assignments. These joint conditions typify probable evolution for a selected quality. Least selection has been quantified in two forms: the largest number of SGC-identical codes evolved at a given time **(51)**, and what was termed plausibility; the largest ratio, (number of SGC-like codes/time for them to evolve) **(1)**.

**Optimized pathways; choosing the best way.** A recurring type of question must be answered: one has a group of events known to be influential; for example, wobble coding. Such events are likely to change either time to evolve or difference from the SGC or both. Thus: what succession of events offers the least selection **(49)**?

To quantitatively respond, a structured list of all possible pathways is made: that is, listed code evolution events occur or do not occur in every possible combination. Time to evolve and distance to the SGC (as well as many other indices, **(51)**) are computed for this comprehensive pathway list. Interesting indices are plotted versus the numbered list as abscissa. Effects of multiple pathway variations can be read off a single 2D plot, appearing as periodicities along the ordinate that reflect the periodicity of evolutionary events in the structured list. For example, in a comparison of 32 pathways with and without simple Crick wobble **(79)** and four other events, wobble hindered code evolution in all 16 cases that allow wobble/no wobble comparison **(51)**. So, in 16 mechanistically-varied contexts, early evolution of wobble is always disfavored **(44)**. Similar multi-dimensional analysis also localized effective code fusion in the RNPT **(1)**.

**Figure legend**

**Figure 1 – Evolution of the Standard Genetic Code**. Dots mark items that are discussed and referenced in the main text. Another account of the same early, code-forming times, from an alternative point of view and with further references, can be found in **(13)**. aa – amino acid; AARS – Aminoacyl-RNA Synthetase; SGC – Standard Genetic Code; LUCA – Last Universal Common Ancestor.

## References

1. M. Yarus, Familiar biological, chemical and physical events credibly evolve the Standard Genetic Code. *arXiv.org* (2024). Available at: https://arxiv.org/abs/2406.08302v1 [Accessed 25 July 2024].

2. W. Gilbert, The RNA world. *Nature* **319**, 618 (1986).

3. K. Kruger, *et al.*, Self-Splicing RNA: Autoexcision and Autocyclization of the Ribosomal RNA Intervening Sequence of Tetrahymena. *Cell* **31**, 147–157 (1982).

4. C. Guerrier-Takada, K. Gardiner, T. Marsh, N. Pace, S. Altman, The RNA Moiety of Ribonuclease P Is the Catalytic Subunit of the Enzyme. *Cell* **35**, 849–857 (1983).

5. M. Yarus, Amino Acids as RNA Ligands: A Direct-RNA-Template Theory for the Code's Origin. *J. Mol. Evol.* **47**, 109–117 (1998).

6. R. M. Turk-Macleod, D. Puthenvedu, I. Majerfeld, M. Yarus, The plausibility of RNA-templated peptides: simultaneous RNA affinity for adjacent peptide side chains. *J Mol Evol* **74**, 217–25 (2012).

7. L. E. Orgel, Evolution of the Genetic Apparatus. *J Mol Biol* **38**, 381–393 (1968).

8. L. E. Orgel, Some consequences of the RNA world hypothesis. *Orig Life Evol Biosph* **33**, 211–8 (2003).

9. M. Yarus, On translation by RNAs alone. *Cold Spring Harb Symp Quant Biol* **66**, 207–215 (2001).

10. M. Yarus, The Genetic Code and RNA-Amino Acid Affinities. *Life* **7**, 13 (2017).

11. R. D. Knight, L. F. Landweber, M. Yarus, "Tests of a stereochemical genetic code" in *Translation Mechanisms*, J. Lapointe, L. Brakier-Gingras, Eds. (Kluwer Academic/Plenum, 2003), pp. 115–128.

12. M. Yarus, Evolution of the Standard Genetic Code. *J. Mol. Evol.* **89**, 19–44 (2021).

13. M. Yarus, From initial RNA encoding to the Standard Genetic Code. [Preprint] (2023). Available at: https://www.biorxiv.org/content/10.1101/2023.11.07.566042v2 [Accessed 12 December 2023].

14. M. Gorlero, *et al.*, Ser-His catalyses the formation of peptides and PNAs. *FEBS Lett.* **583**, 153–156 (2009).

15. R. Wieczorek, M. Dörr, A. Chotera, P. L. Luisi, P.-A. Monnard, Formation of RNA phosphodiester bond by histidine-containing dipeptides. *Chembiochem Eur. J. Chem. Biol.* **14**, 217–223 (2013).

16. K. Adamala, J. W. Szostak, Competition between model protocells driven by an encapsulated catalyst. *Nat. Chem.* **5**, 495–501 (2013).

17. P. Janković, D. Kalafatovic, Short catalytic peptides with tunable activity: Cys confers functionality and adaptability. [Preprint] (2024). Available at: https://chemrxiv.org/engage/chemrxiv/article-details/6626e07591aefa6ce1210b7c [Accessed 9 October 2024].
12

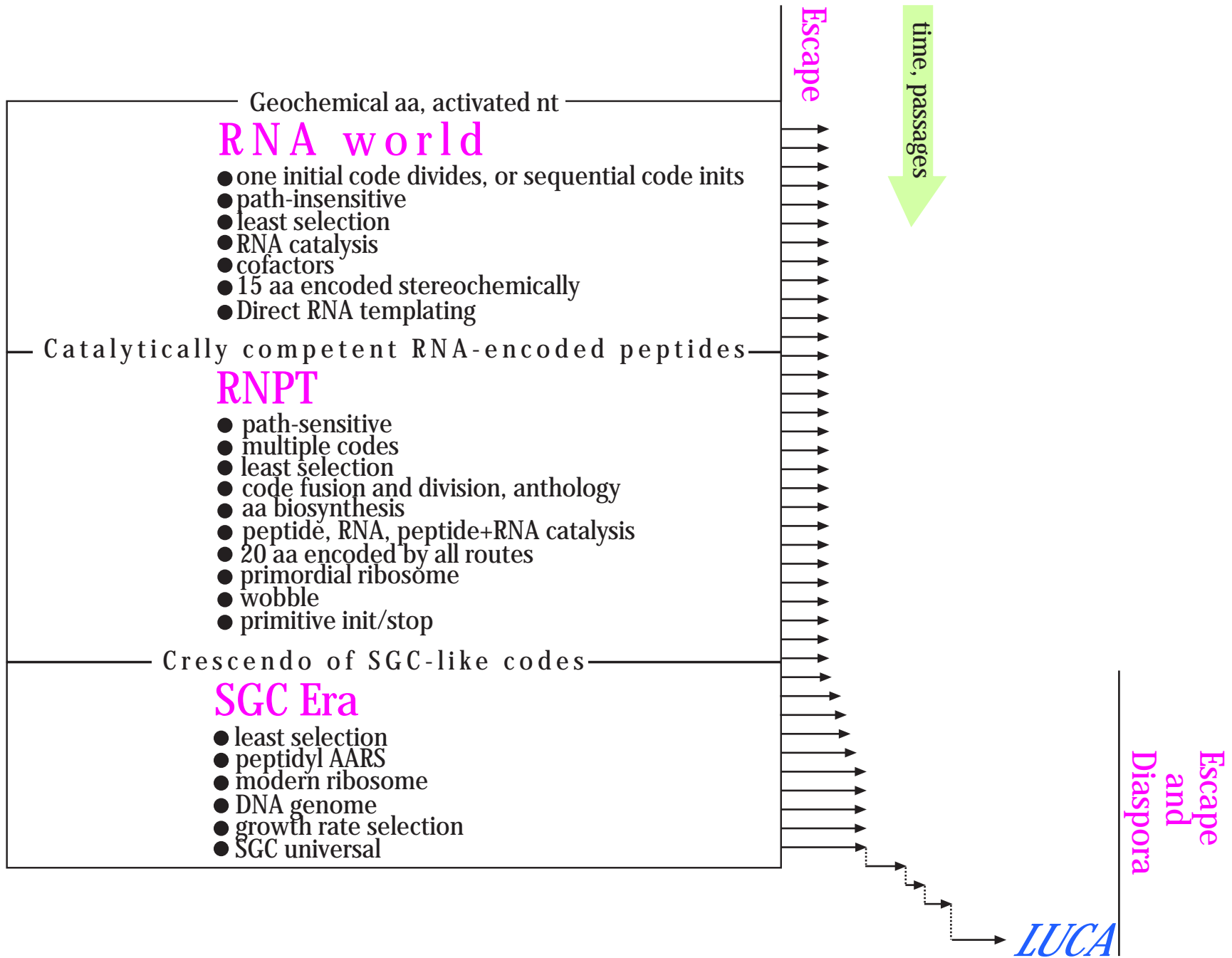

Figure 1